\documentclass[aps,amssymb,prl,twocolumn,showpacs,superscriptaddress]{revtex4}
\pdfoutput=1
\usepackage[english]{babel}
\usepackage[latin1]{inputenc}
\usepackage[dvips]{graphicx}
\usepackage{amsmath, amssymb}
\usepackage{epsfig}
\usepackage{graphicx}
\usepackage{array}

\usepackage{cleveref}
\usepackage{color}

\begin{document}


\title{Hydrodynamic Swarming of Thermally Active Dimeric Colloids}
\author{Martin Wagner\email{m.wagner@fz-juelich.de}} 
\affiliation{Theoretical Soft-Matter and Biophysics,
  Institute of Complex Systems,
  Forschungszentrum J\"ulich, 52425 J\"ulich, Germany}
\author{Marisol~Ripoll\email{m.ripoll@fz-juelich.de}}
\affiliation{Theoretical Soft-Matter and Biophysics,
  Institute of Complex Systems,
  Forschungszentrum J\"ulich, 52425 J\"ulich, Germany}

\date{\today}

\begin{abstract}
  Self-propelled phoretic colloids have recently emerged as a
  promising avenue for the design of artificial swimmers. These
  swimmers combine purely phoretic interactions with intricate
  hydrodynamics which critically depend on the swimmer shape.
  Thermophobic dimer shaped colloids are here investigated by means of
  hydrodynamic simulations, from the single particle motion to their
  collective behavior. The combination of phoretic repulsion with
  hydrodynamic lateral attraction favors the formation of planar
  moving clusters. The resulting hydrodynamic assembly in flattened
  swarms is therefore very specific to these dimeric active colloids.
\end{abstract}

\pacs {66.10.cd,
87.17.Jj,
05.70.Ln,
02.70.Ns}

\maketitle

Synthetic microscale motors are attracting large attention due to
their outstanding potential practical applications in fields like
microfluidics or microsurgery~\cite{paxton04,dreyfus05,elgeti_rev}.
When the propulsion mechanism is based on phoretic
effects~\cite{anderson89}, such artificial microswimmers have the
great advantage of behaving like passive colloids unless they are
chemically~\cite{paxton04,golestanian07,valadares10},
electrically~\cite{loget11,seifert12}, or thermally
activated~\cite{jiang10,yang11dimer,kroy16}.  In particular,
thermophoretic swimmers are built from two materials with well
differentiated absorption coefficients, like gold and silica, where
heterogeneous heating can produce a steady local temperature gradient
around the colloid, which translates into its persistent
self-propulsion. Thermophoretic swimmers can therefore be powered
without any modification of the solvent, what makes them easily
bio-compatible.  Furthermore, devices engineered with this effect are
expected to depict a very large versatility due to two additional
facts. One is that thermophoresis has shown to be very sensitive to a
large number of factors like pressure, average temperature or solvent
composition; and two is that the heat sources, such as magnets or
lasers, can be very precisely controlled in time and space.

Large ensembles of phoretic swimmers are expected to share a large
number of properties with other systems of active particles.
Chemically active Janus colloidal particles have already shown
clustering and self-assembled structures~\cite{kapral13jcp} as well as
schooling behavior; and the formation of living crystals has already
been observed for light powered micromotors~\cite{ibele09,palacci13}.
Brownian simulations of thermophilic active colloids predicted the
appearance of clustering and comet-like swarming
structures~\cite{golestanian12,cohen14prl}.  Nevertheless, the
mechanisms involved in the formation of these structures, the
importance of the phoretic and hydrodynamic effects, and the behavior
of various types of phoretic swimmers are still very relevant and
largely unexplored questions.
Until now, spherical Janus thermophoretic swimmers have been the only
geometry investigated experimentally~\cite{jiang10,cichos13,baraban13}, 
although important effects related with particle shape can be expected.
Catalytic dimer motors, moving in the direction of the catalytic
  cap, have already been synthesized~\cite{valadares10}.  These
results indicate that diffusio- and thermophoretic motion of dimeric
colloids in both directions are experimentally feasible.

In this letter we investigate thermophobic dimer shaped colloids by
means of computer simulations.  Thermophobic colloidal behavior is
much more frequent than thermophilic one. Furthermore, these active
colloids constitute a paradigmatic case in which the interplay of
phoretic repulsion and hydrodynamic attraction results in spontaneous
self-assembly of oriented and collectively moving structures,
  reminiscent of planar order.  These remarkable swarming structures
appear then to be specific to phoretic active colloids; their
  direction of motion will be easier to control~\cite{lozano16}
such that promising applications are to be expected, as those
related to directed cargo transport~\cite{soler13} or the development
of microfluidic devices~\cite{baigl12,16tratchet}. 

Simulations are performed with a hybrid mesoscopic approach.
Multiparticle collision dynamics (MPC) is the particle based method
used to describe the fluid, and molecular dynamics (MD) is employed to
describe the colloids and the colloid-fluid
interactions~\cite{kap99,kapral_review}.  This description has shown
to properly incorporate hydrodynamic interactions, also in phoretic
systems~\cite{kapral09,luese12jcp,yang13flow}.  MPC describes the
fluid as a collection of point particles of mass $m$ that perform
alternating streaming and collision steps.  In the streaming step,
particles propagate ballistically for a time $h$.  In the collision
step, the particles are binned into cells of side length $a$. A grid
shifting procedure~\cite{ihl03a} is employed during this binning to
ensure Galilean invariance.  Inside the collision cells, the particles
velocities relative to the center-of-mass velocity of the cell are
rotated around a random axis by an angle $\alpha$.  The choice of
$a=1,m=1$ and $k_\mathrm{B}\overline{T}=1$ defines the simulation
units, so that time is scaled with
$\sqrt{ma^2/k_\mathrm{B}\overline{T}}$ and velocity with
$\sqrt{k_\mathrm{B}\overline{T}/m}$. The other parameters are here
chosen to be $\alpha=120^\circ$ and $h=0.1$, together with the
averaged number of particles per collision cell, $\rho=10$.  These
numbers determine the fluid transport properties as the solvent
Schmidt number, which here is $Sc=13$, or the fluid
  kinematic viscosity $\nu=0.79$.

Colloid-fluid interactions are modelled with MD, and the choice of the
potential is crucial to determine phoretic and collective properties
of the colloids.  Thermophobic behavior has been observed with
attractive interactions, and thermophilic with repulsive
ones~\cite{luese12jpcm}.  On the other hand, a pronounced depletion
interaction between colloids might occur as result from the high
compressibility of the MPC fluid~\cite{pad06}. A detailed
investigation shows this effect to be very pronounced when using steep
attractive potentials~\cite{wagner3}. The colloid-fluid potential
$U(r)$ and related parameters are then chosen to avoid such artificial 
depletion in the system. We use displaced Lennard-Jones-type potentials given by
\begin{equation}
U(r) = \left\{\begin{tabular}{lc}
$\infty,$ & $r \le \Delta$ \\
$4 \varepsilon \left[ \left( \frac{\sigma}{r - \Delta} \right)^{48} 
- \left( \frac{\sigma}{r - \Delta} \right)^{24} \right] + C,$& 
$\Delta < r < r_\mathrm{c}$\\
$0,$&$ r_\mathrm{c} \le r $
\end{tabular}\right.
\label{lj} 
\end{equation} 
Here $r$ is the pairwise distance, $\varepsilon$ describes the
strength of the potential which we choose as $\varepsilon=k_\mathrm{B}\overline{T}$, and
$\Delta$ introduces a displacement. Repulsive
interactions are obtained with $C=\epsilon$ and
$r_\mathrm{c}=2^{1/24}\sigma+\Delta$, and attractive with $C=0$ and
$r_\mathrm{c}=1.13\sigma+\Delta$.
With this model, the effective radius of each bead is
$s=\sigma+\Delta$, and we denote the related size parameters as
$(s,\Delta)$.  Dimers investigated here are constructed by one
phoretic bead with attractive interactions of size $s_\mathrm{P}$,
which we fix as $(6,3)$; and one hot bead with repulsive interactions
of size $s_\mathrm{H}$, which we mostly fix as either $(6,3)$ or
$(2,0.5)$.  Both beads are held together by a strong harmonic
potential at a distance $s_\mathrm{H}+s_\mathrm{P}$.
Similar to previous works~\cite{yang11dimer,yang14janus}, heating is
modeled by rescaling the temperature of fluid particles in a short
layer ($0.08 s_\mathrm{H}$) around the hot beads to a value of
$T_\mathrm{h}=1.5$, while keeping the overall average fluid
temperature at $\overline{T}=1.0$ using simple velocity rescaling.
This constant heating neglects shadowing effects~\cite{cohen14prl}, it
corresponds to colloids made of materials with the same light
refraction index as the solvent, and which in the case of thermophobic
behavior will have smaller influence. Simulations are performed using
a modified variant of the software package \textsc{lammps}~\cite{Plimpton1995},
in particular a modified version of the
'srd'-package~\cite{Petersen2010}. The time-step to integrate the
potential interactions is $\Delta t = 0.01 h$, and the bead mass is
chosen such that the swimmers are neutrally buoyant.
 
\begin{figure}[h!]
\includegraphics[width=.49\columnwidth]{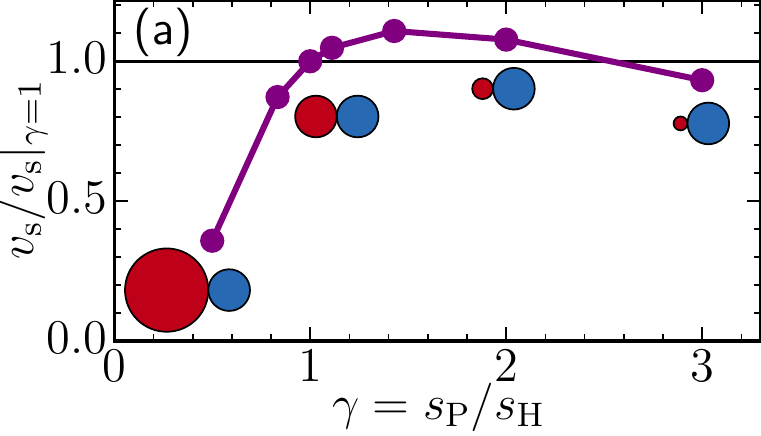}
\includegraphics[width=.49\columnwidth]{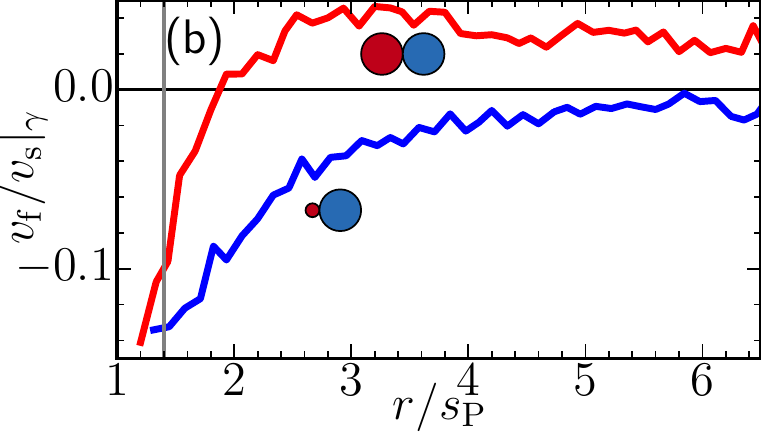}

\includegraphics[width=.49\columnwidth]{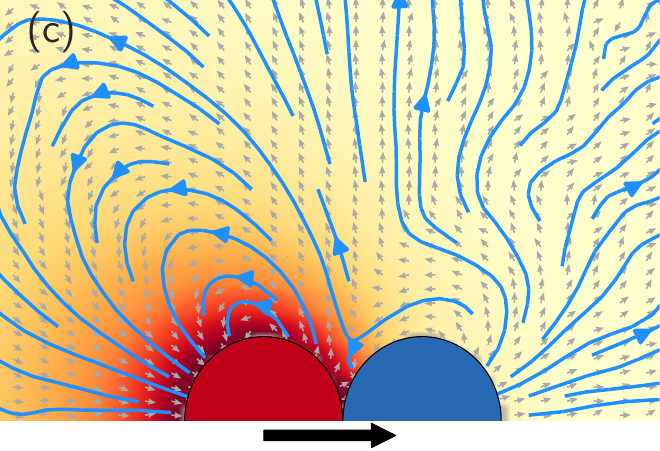}
\includegraphics[width=.49\columnwidth]{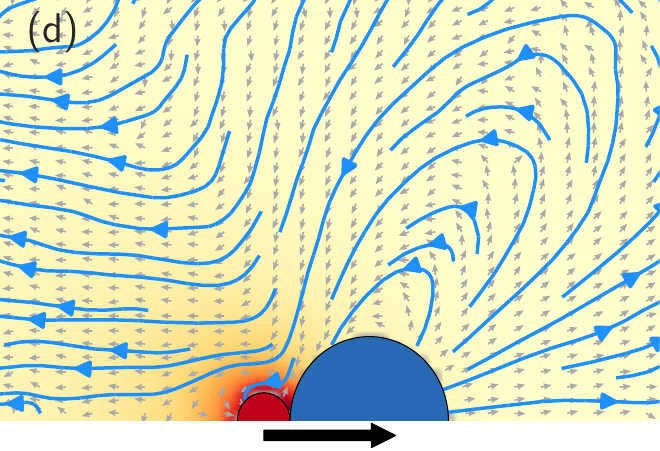}
\caption{a)~Single swimmer velocity $v_\mathrm{s}$ as a function of the bead
  size ratio $\gamma$ normalized by the velocity of the symmetric
  swimmer, $v_s\large|_{\gamma=1}$. 
  b)~Fluid velocity $v_\mathrm{f}$ as a function of $r$, the distance to
  the center of the phoretic bead in the axis perpendicular to the
  swimmer orientation. The normalizing factor $v_\mathrm{s}\large|_{\gamma}$ is
  the corresponding single swimmer velocity. 
  Vertical grey line is the minimum distance from
  the phoretic bead center to the surface of another dimer.  
  Blue (dark) lines stand for symmetric dimers and red (light) for
  asymmetric dimers.
  c)~Flow field characterization in the area close to the symmetric
  dimer $\gamma=1$.  Solid lines correspond to the stream lines,
  background color depicts the temperature field, and grey arrows show
  the direction of the fluid velocity.  Black thick arrow indicate the
  dimer swimming direction. 
  d)~Flow field close to the asymmetric dimer $\gamma=3$, with symbols
  similar to~c). 
\label{single}}
\end{figure}

Besides the employed colloid-fluid interactions, the swimmer
dimensions importantly influence both the swimmer velocity and
hydrodynamic behavior.  Given a constant temperature gradient, the
size of the phoretic bead will determine the phoretic thrust, which
will be larger, the larger the bead~\cite{piazza08,luese12jpcm}.  
In this study the size of the phoretic bead is fixed to
$s_\mathrm{P}=6$.  The symmetric swimmer with
$s_\mathrm{H}=s_\mathrm{P}$ shows a propelled velocity against the
heated bead with an averaged value of $v_\mathrm{s}=0.021$, which
corresponds a Reynolds number $\mathrm{Re}=v_\mathrm{s}s/\nu= 0.32$,
where the relevant particle size is $s=s_\mathrm{H}+s_\mathrm{P}$. At
such low Reynolds numbers, inertial effects can be neglected and
Stokes hydrodynamic behavior is to be expected. 
Increasing the size of the heated bead increases the temperature
gradient around the phoretic bead, such that the phoretic thrust is
larger, but it also increases the overall friction of the swimmer. The
combination of these two effects results in a maximum of the propelled
swimmer velocity as a function of the size ratio $\gamma =
s_\mathrm{P}/s_\mathrm{H}$ of the two dimer beads as can be seen in
Fig.~\ref{single}a. The dimer velocity is measured in the main dimer
axis towards the phoretic bead. The simulations were performed for
single swimmers in a cubic periodic box of side length
$10(s_\mathrm{H}+s_\mathrm{P})$.
The bead size ratio of the dimer also changes its rotational diffusion
coefficient which we have measured for $\gamma=1$ as $D_r=2.5\times
10^{-5}$ and for $\gamma=3$ as $D_r=1.2\times 10^{-4}$, what allows
us to determine the Peclet number $\mathrm{Pe}=v_\mathrm{s}/(s D_r)$ of both
swimmers as $70$ and $20$ respectively. These dimensionless numbers
are comparable, and even a bit higher than those of experimentally
synthesized spherical Janus swimmers~\cite{jiang10,bechinger11}.

The swimmer shape also influences strongly the hydrodynamic response
of the surrounding fluid. %
The quantitative values of the flow velocities in the axis
perpendicular to the swimmer orientations is depicted in
Fig.~\ref{single}b, and the stream lines around the two thermophoretic
dimers are displayed in Figs.~\ref{single}c,d. 
Negative values of the velocity correspond to fluid streaming towards
the bead ({\em hydrodynamic attraction}), while positive values
correspond to the fluid streaming away from the bead ({\em
  hydrodynamic repulsion}). 
The dimers flow fields are of the point force-dipole
type~\cite{popescu11,reigh15}, whose precise form strongly depends on
the geometry of the swimmer. 
The symmetric dimer, with $\gamma=1$, shows mainly hydrodynamic
lateral repulsion, although there is a small region close to the
phoretic bead where attractive interactions also exist.  This
qualitatively changes in the case of asymmetric dimers, with
$\gamma=3$, which show strong and long ranged lateral hydrodynamic
attraction.  This can be seen in Fig.~\ref{single}d, where the stream
lines show a large attractive lobe close to the phoretic bead, and in
Fig.~\ref{single}b where the fluid velocity is clearly negative over
the whole range of accessible distances.

The described hydrodynamic behavior, combined with the thermophobic
character of the dimers, strongly suggests that the collective
dynamics of these dimers will be different from each other and
from other synthetic swimmers~\cite{thakur12,palacci13}. 
To characterize the collective dimer dynamics, we performed
simulations with $N=100$ swimmers at a volume fraction of $\phi=0.05$
for thermophobic swimmers of both size ratios. 
This number of swimmers is limited by the employed large bead
diameters, which implies the use of up to $4 \times 10^7$ MPC
fluid particles. 
Colloid-colloid strong repulsive interactions are considered between
all $i,j$ beads of different dimers, by using potentials corresponding to
Eq.~(\ref{lj}) with $\Delta=0, \varepsilon=2.5
k_\mathrm{B}\overline{T}$, and $C=\varepsilon$. The interaction
distance is $\sigma = 1.2(s_i+s_j)$, which includes a small additional
separation to resolve lubrication forces~\cite{pad06} and to
avoid residual depletion while maintaining the hydrodynamic
interactions.

\begin{figure}[h!]
{\includegraphics[width=.48\columnwidth]{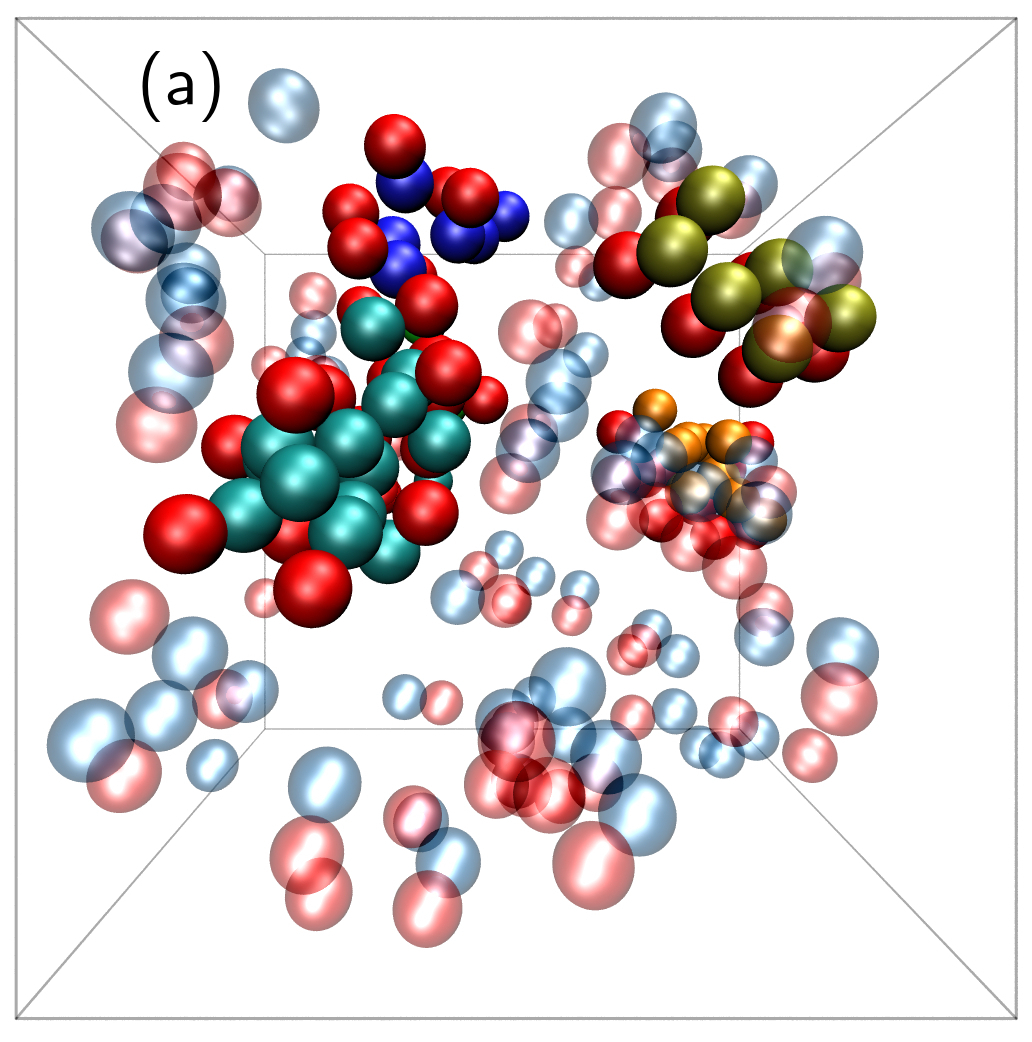}}
{\includegraphics[width=.48\columnwidth]{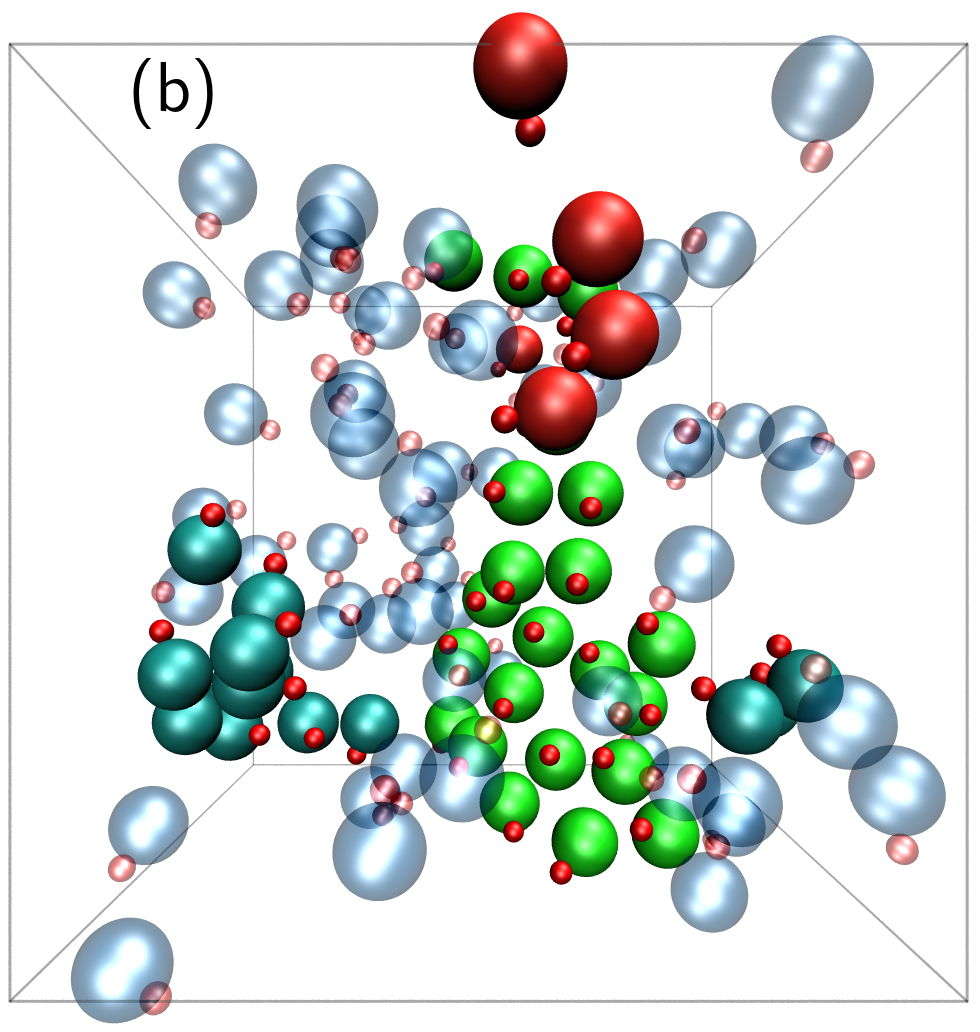}}
\caption{Snapshots of ensembles of thermophoretic dimers.
  a)~Symmetric swimmers ($\gamma=1$).  b)~Asymmetric swimmers
  ($\gamma=3$).  Non-assembled dimers are translucent, with red heated
  beads and blue phoretic beads.  Dimers assembled in clusters of size five
  or larger are solid, with red heated beads and phoretic beads colored
  according to cluster identity. See related movies~\cite{movie}.}
\label{snapshots_collective}
\end{figure}

Thermophobic dimers with the two considered symmetries have a very
dynamic behavior, constantly assembling and de-assembling in clusters
of various sizes. 
Snapshots illustrating different kinds of behavior are shown in
Fig.~\ref{snapshots_collective}, and two representative movies can be
found in~\cite{movie}. In Fig.~\ref{snapshots_collective}a, four
clusters of symmetric dimers can be distinguished, the two blue ones
are randomly jammed structures, resulting from collisions of various
dimers or existing smaller clusters, swimming in random
directions. The other two clusters show clear orientational order,
which will be necessarily accompanied by a significant average
velocity. The light green cluster of asymmetric dimers in
Fig.~\ref{snapshots_collective}b shows a distinct large planar and
oriented structure, with certain hexagonal order, which will
collectively move as a front. 
The strong lateral hydrodynamic attraction favors asymmetric dimers to
swim close and aligned to other dimers, and the phoretic repulsion
avoids the proximity of other dimers in their axial direction,
resulting in the planar structures.
Although the two described assembling mechanisms, hydrodynamics or
collisions, are present for both symmetric and asymmetric swimmers,
the hydrodynamic is clearly more important for the asymmetric ones.
In the case of the symmetric dimers the clusters are small-sized and
short-lived, while asymmetric dimers form considerably larger and
faster clusters with a pronounced tendency to be planar aligned, which
can propel over one to several dimer lengths.  These swarming clusters
are in strong contrast with the observed behavior of spherical
diffusiophoretic colloids in quasi-2D
confinement~\cite{thakur12,palacci13}, or squirmer-type
swimmers~\cite{zoet14} with attractive interactions, which crystallize
in large static clusters. %

\begin{figure}[h!]
  \includegraphics[width=\columnwidth]{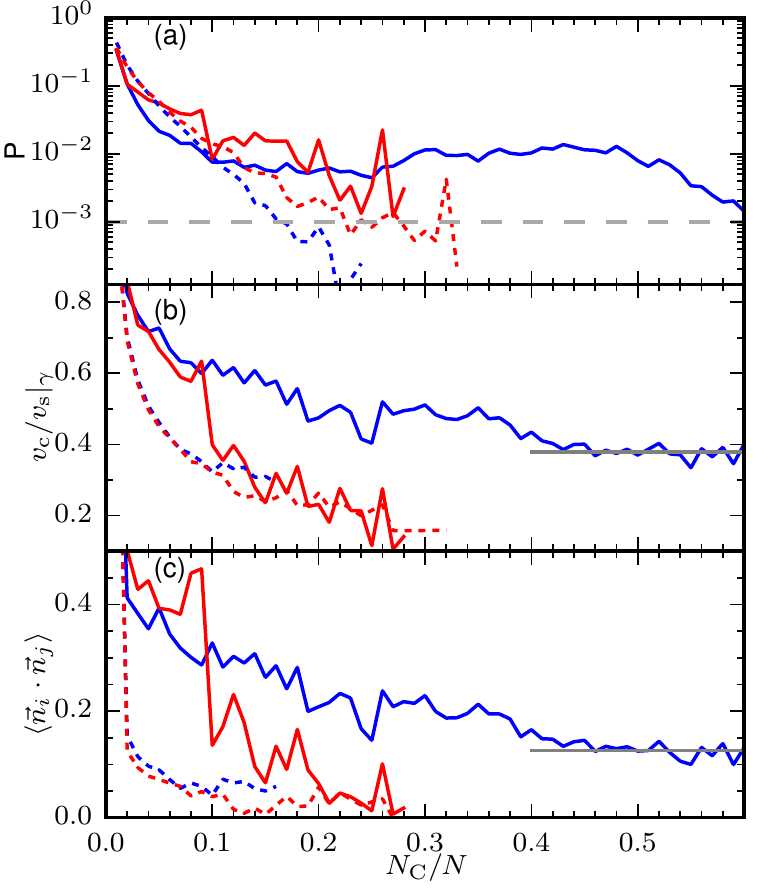} 
  \caption{a)~Averaged probability for a dimer of being in a cluster
    of normalized size $N_\mathrm{C}/N$. 
    Blue (dark) lines stand for symmetric dimers and red (light) for
    asymmetric dimers.  Solid lines correspond to hydrodynamic
    simulations (MPC-MD), dashed lines to non-hydrodynamic results
    (LD).  
    Gray dashed line shows the probability threshold.
    b)~Normalized cluster velocities as a function of normalized
    cluster size. 
    c)~Orientational correlation of dimers within the same cluster as
    a function of the normalized cluster size.  
    b,c) Solid gray lines are a guide to the eye indicating a saturation value.
  }
 \label{fig:analysis}
\end{figure}

To better quantify these effects we perform a cluster analysis of our
simulation results. Any two dimers with beads $i,j$ closer than
$1.32(s_\mathrm{i}+s_\mathrm{j})$ (this is $1.1$ times the
colloid-colloid minimum interaction distance) are considered to belong
to the same cluster. 
Figure~\ref{fig:analysis}a displays the probability of a dimer
to be part of a cluster of size $N_c$. Although most dimers are part
of very small clusters, it can be seen that for symmetric dimers a
significant fraction of them assembles in clusters of up to $\approx
20\%$ of the swimmers; meanwhile asymmetric dimers show to assemble in
significantly larger clusters that might reach over $50\%$ of the
swimmers in the system.  In our discussion, we do not consider data
from cluster sizes with a probability below $10^{-3}$, since they can
be considered rare events.  Note that once a cluster is formed, the
number of constituent dimers is strongly fluctuating, since single
dimers are constantly attaching and detaching, and the cluster is
easily dividing and merging with others. An indication of the most
typical cluster sizes can then be found in the plateaus of the
probability function in Fig.~\ref{fig:analysis}a, which indicates
$N_c=15$ for $\gamma=1$ and $N_c=45$ for $\gamma=3$. 
The cluster velocity along the main cluster direction is plotted in
Fig.~\ref{fig:analysis}b. The main cluster direction is
obtained by averaging all single dimer axes $\vec{n}_i$, and the
cluster velocity is the average of the dimer velocities projected on
such cluster direction. 
The average correlation of all dimer orientations in a cluster
$\langle \vec{n}_i \cdot \vec{n}_j \rangle$ is shown in
Fig.~\ref{fig:analysis}c.  
For symmetric dimers, the velocity and the orientation clearly decay
with cluster size, and although significant propulsion velocity and
orientation is found for small clusters, both are almost vanishing for
the largest clusters of symmetric dimers, as expected for jammed
structures. For asymmetric dimers, velocities and orientations show
similar initial decay, but soon this decay is much slower and it
saturates to a finite value for largest clusters, which is related to
the velocity and the overall orientation of moving planar fronts.

Apart from the hydrodynamic and phoretic interactions, a different
effect present in these systems is an additional inter-dimer alignment
caused by the combination of self-propulsion and steric repulsion that
we here refer to as {\em motility-induced attraction}.  This alignment
occurs since two rod-like propelling particles stay together longer
than non-propelling ones. This effect is similar to an effective
attraction and it has been observed for elongated
swimmers~\cite{abkenar13, Ginelli2010}. 
To evaluate the importance of this effect, we perform Langevin
Dynamics (LD) simulations, where hydrodynamics and phoretic effects
are completely disregarded. Dimeric propelled particles are simulated
by imposing driving forces that relate to swimming velocities
comparable to those in the hydrodynamic simulations.  Fluid friction
is considered by an implicit solvent that we tune to lead to an
effective viscosity very similar to the one in MPC-MD. 
Results are shown in Fig.~\ref{fig:analysis} with dashed lines.
Although some clustering is in fact also found in the LD simulations,
these are mostly driven by simple collision events, and the
average cluster orientation and velocity is clearly much smaller than
in the MPC-MD ones. This proofs that motility-induced attraction
cannot be the cause of the observed swarming behavior. Instead, the
observed swarming structures in the presence of a hydrodynamic
solvent, especially those of the asymmetric dimmers (see
Fig.~\ref{single}c), are due to the precise interplay of hydrodynamic
and phoretic effects in these synthetic structures.

In the case of asymmetric dimeric swimmers where large clusters are
formed, some of our simulations show percolated states in which a
giant cluster extends over the whole simulation box. Upon percolation,
almost all swimmers become part of one big, persistent planar and very
fast front, with only little fluctuations.  The appearance and
stability of this giant front is a consequence of the periodic
character of the employed boundary conditions. This means that it can
be considered a simulation artifact and we have disregarded these
realizations in our previous analysis and discussion.  On the other
hand, percolation has only been observed for asymmetric swimmers in a
hydrodynamic solvent, what also highlights the strength of the lateral
attraction in these swimmers.  Furthermore, it is also important to
note that the maximum size of our observed clusters is strongly
limited by the size of our system; such a limitation does not exist in
experimental systems. It is therefore reasonable to expect that very
large and fast swarming clusters can be soon experimentally observed
in systems of dimeric phoretic colloids.

In conclusion, we have shown that active thermophoretic asymmetric
dimers assemble in moving clusters with flattened structures resulting
from the interplay of hydrodynamics and phoresis. Although some
swarming is also observed in symmetric thermophobic dimers, the
stronger and longer ranged hydrodynamic interactions of the asymmetric
active dimers makes that their clusters are considerably larger,
faster, and more coherently oriented. 
This particular type of swarming can also be expected in other
phoretic structures, such as catalytic dimeric swimmers with motion
towards the non-catalytic surface. 
On the other hand, this behavior stands in contrast to most other
artificial active systems based on spherical particles as well as to
phoretically attractive systems, where activity leads to clustering
into unmoving structures. In systems of biological active particles,
various ways of swarming have been described, although, to our
knowledge, not in planar fronts as those here described. 
This novel phenomenology will broaden the scope of possible
applications of active matter and offers intriguing new possibilities,
for example in the design of microfluidic devices, or bio-compatible
micromotors.

\vspace*{0.5cm}
This work was supported by the DFG priority program SPP 1726 on
``Microswimmers - from Single Particle Motion to Collective Behaviour''.
The authors also gratefully acknowledge the computing time granted on
the supercomputer JURECA at J{\"u}lich Supercomputing Centre (JSC).


\end{document}